# Damped Lyman Alpha Systems vs. Cold + Hot Dark Matter

**Anatoly Klypin** (aklypin@nmsu.edu)
Department of Astronomy, New Mexico State University
Las Cruces, NM 88001

**Stefano Borgani** (borgani@vaxpg.pg.infn.it)
INFN, Sezione di Perugia, c/o Dp. di Fisica dell'Università
via A. Pascoli, I-06100 Perugia, Italy
and
International School for Advanced Studies
Via Beirut 2-4, I–34014 Trieste, Italy

**Jon Holtzman** (holtz@lowell.edu)
Lowell Observatory, Mars Hill Road, Flagstaff, AZ 86100

**Joel Primack** (joel@lick.ucsc.edu)
Physics Department, University of California, Santa Cruz, CA 95064

**Abstract** Although the Cold + Hot Dark Matter (CHDM) cosmology provides perhaps the best fit of any model to all the available data at the current epoch ($z = 0$), CHDM produces structure at relatively low redshifts and thus is very sensitive to the observed numbers of massive objects at high redshifts. Damped Ly$\alpha$ systems are abundant in quasar absorption spectra and provide possibly the most significant evidence for early structure formation, and thus a stringent constraint on CHDM. Using the numbers of halos in N-body simulations to normalize Press-Schechter estimates of the number densities of protogalaxies as a function of redshift, we find that CHDM with $\Omega_c/\Omega_\nu/\Omega_b = 0.6/0.3/0.1$ is compatible with the damped Ly$\alpha$ data only at $\leq 2.5$, but that it is probably incompatible with the $z > 3$ damped Ly$\alpha$ data. The situation is uncertain because there is very little data for $z > 3$. The predictions of CHDM are quite sensitive to the hot (neutrino) fraction, and we find that $\Omega_c/\Omega_\nu/\Omega_b = 0.725/0.20/0.075$ (and possibly even $\Omega_c/\Omega_\nu/\Omega_b = 0.675/0.25/0.075$) is compatible with the $z > 3$ data. With one massive neutrino species, using $\Omega_\nu = 0.20$ instead of 0.30 corresponds to lowering the neutrino mass from 7.0 to 4.7 eV, for $H_0 = 50$ km s$^{-1}$ Mpc$^{-1}$ and $T = 2.726$ K. In CHDM, the higher redshift damped Ly$\alpha$ systems are predicted to have lower masses ($\sim 3 \cdot 10^{10}$ $M_\odot$ at $z = 3$), a prediction which can be checked by measuring the velocity widths of the associated metal line systems.

Predictions for high-z objects crucially depend on the effects of limited resolution and the finite box size in N-body simulations or on the parameters of the Press-Schechter approximation, if it is used. By analysing our numerical simulations with vastly different resolutions and box sizes as well as those of Ma & Bertchinger (1994), we show that for the CHDM models with $\Omega_\nu$=0.2–0.3 the Press-Schechter approximation should be used with Gaussian filter with $\delta_c = 1.5$ if halos are defined with the mean overdensity larger then 200. If one tries to recover the total mass of a collapsed halo, a better value for the collapse parameter is $\delta_c = 1.40$. We argue that nonlinear effects due to waves both longer and shorter than those considered in numerical simulations could probably result in $\delta_c$ as low as $\delta_c = 1.3$.

*Subject headings:* cosmology: theory — dark matter — large-scale structure of universe — galaxies: formation — quasars: absorption lines

## 1. Introduction

The standard Cold Dark Matter (CDM) cosmology is based on the assumptions that structure grew by gravitational collapse from initial adiabatic Gaussian fluctuations with a Harrison-Zel'dovich spectrum in a critical-density (i.e. $\Omega = 1$) universe as predicted by simple cosmic inflation models, together with the assumption that the dark matter is "cold" – i.e., fluctuations of all cosmologically interesting sizes are preserved during the radiation era (Blumenthal et al. 1984). Although CDM with bias parameter $b \approx 2$ was found to be consistent with data from galaxy to cluster scales, it appears to be less compatible with data on larger scales (see e.g. Davis et al. 1992). In particular, if cosmological theories are normalized to produce cosmic microwave background temperature anisotropies compatible with the COBE measurements (Smoot et al. 1992, Bennett et al. 1994), then Cold Dark Matter (CDM) produces somewhat too much structure on scales of several 10's of Mpc (e.g., too many clusters; cf. White, Efstathiou, & Frenk 1993a) and far too much structure on scales of a few Mpc (e.g., median rms velocities in groups are far higher than observations; cf. Nolthenius, Klypin, & Primack 1994 [NKP94]).

Cold + Hot Dark Matter (CHDM) is based on exactly the same assumptions as standard CDM, except that in addition to the cold dark matter it is assumed that one of the three known species of neutrinos has a mass of $23.5\Omega_\nu h_{50}^2$ eV, where $\Omega_\nu$ is the fraction of critical density contributed by these neutrinos and $h_{50}$ is the Hubble parameter in units of 50 km s$^{-1}$ Mpc$^{-1}$. These neutrinos would have been highly relativistic at the era (approximately a year after the Big Bang) when the horizon first encompassed $10^{12}$ $M_\odot$, since the temperature then was $\sim$keV; thus they would be "hot" dark matter. Their free streaming would have erased fluctuations in their density on scales from galaxies through clusters during the radiation era, and their still-significant random velocities at more recent epochs would have impeded their clustering on small scales. Consequently, if the CDM and CHDM density fluctuation spectra $P(k)$ are both normalized to agree with COBE, as in Figure 1(a), then CHDM has a little less power on intermediate scales and much less power on small scales (i.e., large $k = 2\pi/\lambda$). Approximate linear calculations (Holtzman 1989, Holtzman & Primack 1993; cf. also Schaefer, Shafi, & Stecker 1989; van Dalen & Schaefer 1992; Schaefer & Shafi 1992, 1994; Taylor & Rowan-Robinson 1992; Wright et al. 1992) showed that the most promising parameters for CHDM were $\Omega_c/\Omega_\nu/\Omega_b \approx 0.6/0.3/0.1$. Cosmological N-body simulations have confirmed that CHDM with this combination of parameters appears to be compatible with all the available data at the current epoch (Klypin et al. 1993 [KHPR93] and references therein;). In disagreement with claims of Cen & Ostriker 1994, NKP94 show that velocities predicted by CHDM are compatible with observations on $\sim 1$ $h^{-1}$ Mpc scales [where as usual $h$ is the Hubble parameter in units of 100 km s$^{-1}$ Mpc$^{-1}$]; SB and Klypin & Rhee (1994), Jing & Fang (1994), and Walter & Klypin (1994) show that CHDM predictions are compatible with observed cluster properties; Plionis et al. (1994) show that CHDM reproduces cluster correlations significantly better than several other variants of the CDM model. The same conclusion concerning X-ray properties of clusters predicted by CHDM was reached by Bryan *et al.* (1994), based on N-body + hydrodynamic simulations.

It should be understood that the linear calculations of Holtzman (1989) were done for a fairly widely spaced grid of values of parameters such as $\Omega_\nu$, and that better fits could possibly be found for intermediate values. Holtzman, Klypin, & Primack (1994) have done a new set of linear calculations, including many intermediate values of parameters (some examples are given in Figure 1(a)), in order to help to choose parameters for new high-resolution N-body simulations. Note that for long waves ($k < 1h$ Mpc $^{-1}$) the difference in power between the standard 0.6/0.3/0.1 CHDM model and these variants is small, which insures that all large scale properties (e.g. bulk velocities, number of clusters, and the cluster-cluster correlation function) will be almost unchanged. As we go to smaller scales the difference becomes larger and can be more than 50% on galactic scales. This is already enough to increase significantly the number of galaxies at high redshifts. Another parameter that can affect the spectrum of fluctuations is $\Omega_b$, the contribution of baryons to the mean density. Two sorts of observations tend to pull it in opposite directions. Abundances of light nuclides together with standard Big Bang Nucleosynthesis (BBN) give low estimates: $\Omega_b \sim 0.05$, with $\Omega_b \sim 0.1$ being marginally consistent with the latest observational results (T. P. Walker and D. N. Schramm, private communications, 1994). At the same time, results from X-ray studies of galaxy clusters favor higher values of $\Omega_b$ (White et al. 1993b). Since we think that estimates of the masses of different components of clusters are still not as reliable as the constraints from BBN, we choose $\Omega_b$ higher than the usual BBN value of $0.05h_{50}^2$, but still compatible with BBN. Note that lowering $\Omega_b$ at constant $\Omega_\nu$ increases the amplitude of fluctuations at small scales while the amplitude at large scales is unaffected.

In this paper, we consider the implications of CHDM for damped Ly$\alpha$ systems. Damped Ly$\alpha$ systems are observed as wide absorption troughs in quasar spectra, indicating that the line of sight to the quasar intersects a cloud of neutral hydrogen with a column density $\geq 10^{20}$ cm$^{-2}$ (for a recent review, see Wolfe 1993). Damped Ly$\alpha$ systems are sufficiently common that their abundance can be characterized statistically out to $z \approx 3$ (Lanzetta 1993, Lanzetta et al. 1993 [LWT93]). They thus provide potentially the most reliable and important information presently available on early structure formation. Three recent papers (Subramanian & Padmanabhan 1994, Mo & Miralda-Escude 1994, Kauffmann & Charlot 1994, and Ma & Bertschinger 1994 [MB94]) have claimed that the statistics on damped Ly$\alpha$ systems are strongly incompatible with the predictions of CHDM.

There are some arguments suggesting that the damped Lyα systems correspond to massive protogalaxies. We note that the arguments typically have the sense of double negative statements: observational data are *not incompatible* with large spiral disks. For example, Briggs et al. (1989) found a high column density absorber at $z = 2.04$ with a complicated substructure: at least two or three components in projection; two components in the absorption spectrum with difference in velocities $\Delta v \sim 16$ km s$^{-1}$ and width $\sigma \sim 10$ km s$^{-1}$. These small velocities could be interpreted in favor of a small mass of the system. But Briggs et al. argue that this is likely a large galaxy with rotational velocity $v_{\rm los} \sim 200$ km s$^{-1}$ because a *spherical* model is incompatible with the data. We note that this is an interpretation of the results, which probably excludes only a very naive model. No direct evidence of large velocities was actually found. Thus, observationally, the situation is quite unclear, especially at the higher redshifts $z \gtrsim 3$ that most significantly constrain CHDM. Taking into account the uncertainties in both the interpretation of the observational results and in theoretical predictions about how galaxy progenitors should look at high redshifts, we will consider the possibility that damped Lyα clouds at higher redshifts are protogalaxies less massive than present day $L_*$ galaxies. In §2 we discuss the number density of dark matter halos of various masses as a function of redshift in CHDM, using N-body simulations to normalize Press-Schechter estimates. In §3, we compare the numbers of damped Lyα systems predicted by CHDM to the data of LWT93. §4 presents our conclusions – which are partly in agreement and partly in disagreement with the recent papers just mentioned. We agree that CHDM with $\Omega_\nu = 0.3$ is incompatible with the $z \gtrsim 3$ damped Lyα data, but we show that CHDM with $\Omega_\nu \approx 0.2$ is compatible with this data. To paraphrase Mark Twain's cable to the Associated Press on reading his obituary in 1897: "The reports of my [model's] death are greatly exaggerated."

## 2. Abundance of Dark Halos

The Press-Schechter approximation (Press & Schechter 1974, Efstathiou & Rees 1988, Bond et al. 1992) is commonly used for estimates of the number of halos at different redshifts. For a filter $W$ with radius $r_f$ and mass $M = \alpha_m \rho_0 r_f^3$ ($\rho_0$ being the critical density at present) the Press-Schechter approximation for the number density of halos with mass larger than $M$ at redshift $z$ is

$$N(>M, z) = \int_M^\infty n(m,z)dm = \sqrt{\frac{2}{\pi}} \frac{\delta_c}{\alpha_m} \int_{r_f}^\infty \frac{\epsilon(r_f, z)}{\sigma(r_f, z)} \exp\left(\frac{-\delta_c^2}{2\sigma^2(r_f, z)}\right) \frac{dr_f}{r_f^2} , \quad (1)$$

where

$$\epsilon(r_f, z) = \int k^4 P(k,z) \frac{W(kr_f)}{kr_f} \frac{dW(kr_f)}{d(kr_f)} dk \bigg/ \int k^2 P(k,z) W^2(kr_f) dk ,$$
$$\sigma^2(r_f, z) = \frac{1}{2\pi^2} \int k^2 P(k,z) W^2(kr_f) dk ,$$
(2)

and $P(k,z) = [(\Omega_c + \Omega_b)\sqrt{P_c} + \Omega_\nu\sqrt{P_\nu}]^2$ is the power spectrum of the total density. For a Gaussian filter $W = \exp[-(kr_f)^2/2]$ it is $\alpha_m = (2\pi)^{3/2}$. For a top-hat filter, $\alpha_m = 4\pi/3$.

The Press-Schechter approximation is commonly considered to be a "good approximation." But in detail, it depends on how one uses it (e.g., Gaussian filter or top-hat filter) and what kinds of objects are considered. The approximation becomes very sensitive to the details once rare objects with a small fraction of the mass in them are considered. The reason is simple – we deal with the tail of a Gaussian distribution, where a small change in $\sigma$ or $\delta_c$ produces large variations in results. Different people give different values for $\delta_c$. For example, for the CDM model and Gaussian filter Efstathiou & Rees (1988) gave $\delta_c = 1.33$; Carlberg & Couchman (1989) and Gelb (1992) gave $\delta_c = 1.44$. In principle, results can be different for different cosmological models (e.g. Ma & Bertchinger (1994).

We regard $\delta_c$ as a parameter to be fit to the numbers of halos actually found in the simulations. From different simulations we find an indication for a trend of $\delta_c$ with mass and redshift. While results for present-day clusters in simulations are well described by $\delta_c = 1.5$ (e.g., Klypin & Rhee (1994) and Bryan *et al.* (1994)), small-mass halos at high redshifts are better approximated by slightly smaller value: $\delta_c = 1.40$. This might indicate that one needs to go beyond simple all-scales all-redshifts PS formalism in order to obtain more accurate estimates. A number of analytical investigations show that standard PS approximation with *top-hat* filter and $\delta_c = 1.68$ underestimates the number of collapsed halos at any redshift. Jain & Bertschinger (1993) included second–order effects in the Eulerian evolution of density fluctuations. They found that such effects provide an enhancement of the redshift at which galaxy mass scales go non–linear. Based on the Lagrangian approach provided by the Zel'dovich approximation, Monaco (1994) tried to account for effects of non–spherical collapse in the final shape of the mass function. He found that the resulting mass function can be fitted with the PS expression with $\delta_c \simeq 1.4$. Jedamzik (1994) attempted to deal with the cloud–in–cloud problem in the PS formalism. He concluded that, while the number of high mass ($\sim 10^{15}M_\odot$) objects is overestimated in the standard PS approach, the number of objects on galaxy scales ($10^{10}$–$10^{12}M_\odot$) can be underestimated up to an order of magnitude. This turns into a steepening of the mass function, i.e. into a decrease of the $\delta_c$ parameter.

Unfortunately, in some cases it is not enough just to use results of numerical simulations. Numerical simulations can underestimate the numbers of dark halos due to two effects: small size of the computational volume and the lack of mass resolution. Both effects should be taken into account when normalizing the Press-Schechter approximation.

*2.1 Numerical Simulations and Halo Identification*

In this paper we use results of the following simulations to estimate the numbers of dark halos and to normalize the Press-Schechter approximation:

(i) We use KHPR93 results for simulations done with 25 $h^{-1}$ Mpc box size, PM code with $256^3$ mesh, $128^3$ cold particles and $6 \times 128^3$ hot particles. Halos in the simulations were identified as local maxima of the total density (hot + cold) on a mesh with $97.5h^{-1}$kpc cell size. The mass assigned to a halo was defined as the sum of masses inside the central cell and its nearest 26 neighbors (thus, a $3^3$ cube in space). The effective comoving radius is $0.181h^{-1}$Mpc, which gives proper radius $60h^{-1}$kpc at $z = 2$. The halos should have central overdensity of more than 50 to be considered as candidates for "galaxies". Because a halo is typically split between four or more cells, the limit roughly corresponds to a real overdensity $\approx 200$ (see below).

(ii) Simulations of Klypin, Nolthenius, & Primack (1994, NKP94): 50 $h^{-1}$ Mpc box size, PM code with $512^3$ mesh [the same resolution as in (i)], $256^3$ cold particles and $2 \times 256^3$ hot particles, and the same halo identification as in (i). In order to check whether the actual overdensity used with algorithm was large enough and to gain further insight into the structure of the halos, we applied a more accurate algorithm to our 50 $h^{-1}$ Mpc CHDM$_2$ simulation (NKP94) at $z = 2.2$, and compared its results with our faster maxima-on-mesh approach. The new algorithm finds positions of spheres of given constant radius and gives the mass inside the spheres. The algorithm is an iterative procedure. At the beginning, it places a sphere of given radius at each maximum of density defined on the mesh. Then it finds the center of mass of all cold particles inside the sphere. Then the center of the sphere is displaced to the center of mass. This procedure is then repeated. The algorithm stops once either the displacement is smaller than 0.005 of the cell size or the mass inside the sphere stops rising. The comoving radius of the sphere was chosen to be $60h^{-1}$kpc or 0.61 of the cell size. Note that except for the initial location, this method does not involve the mesh (so there are no problems with the positions of halos relative to the mesh). Typically 4–6 iterations (with a maximum of 10-12) are needed to find the position that corresponds to a local maximum of cold mass inside the sphere.

Finally, we place three spheres of different radii centered on each halo found. The radii were different by a factor 2 from one another: $180h^{-1}$kpc (the effective radius of our maxima-on-mesh approach), $90h^{-1}$kpc, and $45h^{-1}$kpc. The diameter of the smallest sphere is about a cell size, which is marginally resolved. We found the following: (1) The mean density contrast of cold dark matter within the sphere increases by a factor 2–5 when its radius decreases by a factor of two. This is roughly consistent with the factor of four expected for an isothermal distribution with $\rho \propto r^{-2}$. (2) The mass within $180h^{-1}$kpc tightly correlates with the mass found with the maxima-on-mesh approach ($3^3$ cube). The former is 20–25% higher and the spread is 10–20%. The difference is small and could be interpreted as if the effective radius for maxima-on-mesh should be $167h^{-1}$kpc, not $180h^{-1}$kpc. The difference is not very significant. Because of the steep growth of the density toward the centers of the dark halos in the simulations, many halos have density contrasts well above 178 expected for the top-hat model. At $z = 2.2$ there were 863 halos with mass larger than $1.5 \times 10^{11}$ $h^{-1}M_\odot$ and central mesh-defined overdensity larger than 50, which have actual overdensity larger than 178 for any of the radii defined above. Our maxima-on-mesh approach gave 775 halos, or 11% fewer. The difference is mainly due to the 20-25% difference in masses. We conclude that our estimates of number of halos are reasonably accurate. If anything, our simpler algorithm probably *underestimates* the number density of high-z objects.

(iii) Because the finite force resolution in our simulations can affect the predicted numbers of dark halos, we ran another simulation with significantly better resolution. The simulation was done for the $\Omega_c/\Omega_\nu/\Omega_b = 0.675/0.25/0.075$ CHDM model. There were $256^3$ cold and $2 \times 256^3$ hot particles. The mesh was $800^3$, which for our box size $20h^{-1}$Mpc gives formal resolution of $25h^{-1}$kpc in comoving coordinates or $6.2h^{-1}$kpc in proper units at $z = 3$. While the resolution is adequate for the problem, the box size is definitely small to accomodate long waves. This should not severely affect our estimates because at redshift $z = 3$ the amplitude of long waves is small and their contribution can be taken into account using a combination of numerical results and the Press-Schechter approximation. In this case we decided to define our halos in a more traditional way – as objects with mean overdensity 200 inside a sphere. Two other presciptions were also used. One prescription checks whether the overdensity at some radius is larger than 200, but then estimates mass for radius with the mean overdensity 100. Another prescription takes radius corresponding to mean overdensity 100, but additionally checks whether the radius is smaller than 43 kpc (proper radius, $h = 0.5$) and chooses the smaller of the two radii.

(iv) Results of Ma & Bertchinger (1994[MB94]) simulations: 50 $h^{-1}$ Mpc box size, $128^3$ cold particles, P$^3$M code with the force resolution of $25h^{-1}$kpc. Halos were identified for a radius with mean overdensity 200.

The way halos are found definitely affects the results. There are a number of problems to take into account when constructing a halo identification algorithm. One of the concerns is that halos that actually have not yet collapsed will make their way to the "catalog". Traditionally, a threshold of overdensity 200 is taken to get rid of "nonreal" halos. The prescription is motivated by the spherical top-hat model of nonlinear collapse. But even within the framework of the top-hat model the prescription is not correct. The top-hat model says that *characteristic* radius of halo reduces by two as compared with the radius at turn-around point. It does not say that the mass within a radius with the overdensity 200 (or 178, to be more precise) is the total mass of the collapsed object. Note that after the collapse, the object no longer has a homogeneous distribution: it is not top-hat. Numerical simulations of the collapse of an initially homogeneous sphere or ellipsoid indicate that the final density profile can be approximated as

$$\rho(r) = \frac{\rho_0}{(1 + (r/r_c)^2)^{n/2}}, \tag{3}$$

where the slope $n$ is about 4 (e.g., Peebles 1970, Gott 1975, Klypin 1980). The steeper the slope, the closer the final distribution is to the homogeneous case (less mass in the tail). Let's take the steepest slope $n = 4$ (a less steep slope will only strengthen our conclusions) and repeat the usual top-hat model derivation (the conservation of energy and the virial equilibrium at the end). The potential energy and the mass as the function of radius for the system are:

$$\begin{aligned} W_{\rm pot} &= \frac{1}{2\pi} \frac{GM_{\rm tot}^2}{r_c}, \\ M(x) &= 2\pi\rho_0 r_c^3 \left[\arctan(x) - \frac{x}{1+x^2}\right] \to M_{\rm tot} = \pi^2 \rho_0 r_c^3, \quad x \equiv r/r_c \end{aligned} \tag{4}$$

This gives us the core radius $r_c$:

$$E_{\rm in} \equiv -\frac{3}{5}\frac{GM_{\rm tot}^2}{R_{\rm in}} = \frac{1}{2}W_{\rm pot} \Longrightarrow r_c = \frac{5}{12\pi}R_{\rm in} \approx R_{\rm in}/7.54 \tag{5}$$

The value of the central density is fixed by the total mass $M_{\rm tot} = 4\pi/3\rho_{\rm in}R_{\rm in}^3$, where $\rho_{\rm in}$ and $R_{\rm in}$ are initial (turn-around) density and radius: $\rho_0 = (1/5)(48\pi/5)^2 \rho_{\rm in}$. It is convenient to express it in units of the background density at the moment of collapse $t_{\rm coll} = 2t_{\rm in}$:

$$(\rho_0/\rho_b)|_{\rm coll} = \frac{1}{5}\left(\frac{72\pi^2}{5}\right)^2 \approx 4040 \tag{6}$$

We can estimate the fraction of collapsed mass within the radius which corresponds to the mean overdensity 200. The fraction is 0.61 and radius is $3.05 r_c$. Thus, almost half of the mass would be missing if we take this limit. If we would like to recover 3/4 of the mass, the radius should be $5.0 r_c$ and the mean overdensity is 58.8.

What actually should be taken as a criterion from the top-hat model is the limit on the *central* density: if the central overdensity in the simulation is less then 200, the object should not be considered (within the framework of spherical top-hat collapse) as collapsed. But if the central overdensity is larger than 200 (ensuring that we deal with a collapsed object), its radius should be larger (within a factor of two for realistic estimates) than the radius corresponding to the mean overdensity 200.

*2.2 Results of Numerical Simulations and Parameters of the Press-Schechter Approximation*

In Figure 1(b) we compare the numbers of halos of masses $1.5 \times 10^{11}$, $5 \times 10^{11}$, and $1.5 \times 10^{12}$ $h^{-1} M_\odot$ from the KHPR93 (i) and NKP94 simulations (ii) (shown as dashed and solid lines, respectively) with the Press-Schechter estimates (dot-dashed lines).

Figure 2 presents our results for the (iii) simulation (20 $h^{-1}$ Mpc box, $800^3$ mesh). The top panel shows the evolution of the number density of halos. The bottom panel compares numerical results at $z = 3.2$ with the Press-Schechter approximation. The low triangles are the predictions of the Press-Schechter approximation with the Gaussian filter, $\delta_c = 1.5$, and the power spectrum actually considered in the simulation – only waves shorter than the box size and longer than the Nyquist frequency of particles were considered. The full curve is for the halos with overdensity 200. The Press-Schechter approximation gives a quite reasonable fit for masses less than $5 \times 10^{10}$ $h^{-1} M_\odot$ . For larger masses P-S falls short by a factor of two, but the number of halos with this mass is small in the simulations ($\approx 10$).

Motivated by the mass estimates in the top-hat model, we tried two other prescriptions. The dot-dashed curve in Figure 2 corresponds to halos whose central overdensity is larger than 200, but whose mass is estimated at the overdensity 100. In this case, the algorithm finds the same halos as before, but assigns larger masses to them (typically by a factor 1.5). We also tried one more prescription, which is close to one we used for our low resolution simulations. We took the overdensity threshold 100, but put an additional constraint on the radius of halos (to exclude too puffy ones): the proper radius should be less than 43 kpc ($h = 0.5$), which corresponds to 3.6 simulation cells and to comoving radius 180 kpc. Note that in this case large halos are assigned smaller masses because some of them had radius bigger then 43 kpc. Smaller halos were the same as found with the additional check of the central overdensity. Deviations were found only for very small halos (less than $2 \times 10^9$ $h^{-1} M_\odot$ ). This means that all our halos with mean overdensity 100 had central overdensities larger than 200. The parameter for the Press-Schechter approximation was $\delta_c = 1.4$ for these two prescriptions.

Comparing to the KHPR93 (i) and NKP94 (ii) simulations, we found that there is no unique value for the threshold parameter $\delta_c$ which can provide a good fit for all masses. The value of $\delta_c$ for a Gaussian filter for the Press-Schechter curve drawn for each mass is indicated in Figure 1(b). When finding dark halos in numerical simulations, we used a constant comoving sphere with effective radius of $0.18 h^{-1}$ Mpc. This is probably small for *large* objects with mass $\geq 10^{12}$ $M_\odot$ . As the result, we miss a number of objects with this mass, which at least partially explains why the numerical results go below the Press-Schechter curves. But the objects that we missed are probably small groups of galaxies, not isolated galaxies. The large value of $\delta_c = 1.60$ found by KHPR93 was mainly based on simulations with very small box size $7h^{-1}$Mpc. Such a small box does not have large waves to produce large structures like superclusters or filaments, which should host high-redshift galaxies. Thus, it is quite likely that results from small boxes significantly underestimate the number of high redshift galaxies. Similarly, note that in Figure 1(b) the number densities from the smaller box are somewhat lower at higher redshift, although they agree at low redshift. This suggests that still larger boxes would give even higher number densities at high redshift. Predictions for Gaussian filter with $\delta_c = 1.33$ and $M = 1.5 \times 10^{11}$ $h^{-1}M_\odot$ are shown as the dotted curve. This curve goes above the results for the large box, but at $z > 2$ the difference is about the same as the difference between the results for small ($25h^{-1}$Mpc) and large ($50h^{-1}$Mpc) box simulations. It is quite reasonable to suggest that the numbers of halos with even larger boxes and better resolution would be close to $\delta_c = 1.33$ predictions. This is also supported by our more accurate estimates of numbers of dark halos at $z = 2.2$. For the comparisons with data on damped Ly$\alpha$ systems in the next section, we therefore use $\delta_c = 1.33$, the same value used by Efstathiou & Rees (1988) for CDM. The top-hat model with traditional $\delta_c = 1.68$ (triangles) predicts systematically smaller numbers of dark dark halos at all redshifts. At $z < 1.5$ the difference, though visible, is not large. But it grows as we go to higher $z$. By $z = 3$ the top-hat model underestimates the number of halos by a factor of four compared to the results from the large-box simulation.

KHPR93 noted that the $\Omega_c/\Omega_\nu/\Omega_b = 0.6/0.3/0.1$ CHDM model predicted almost as many large galaxies at high redshifts as CDM with $b = 2.5$, probably enough to account for the numbers of bright quasars. The conclusions of Haehnelt (1993), Liddle & Lyth (1993), and Pogosyan & Starobinsky (1993) are that this model may predict too few quasars if the number density of bright quasars at $z \approx 4$ is similar to that at $z = 2 - 3$. However, Warren, Hewett, & Osmer (1994) find a steep drop-off in the number density for redshifts beyond about 3.3, so even the 0.6/0.3/0.1 CHDM model may have no problem accounting for number densities of bright quasars.

Recently Ma & Bertchinger (1994) claimed, on the basis of their simulations (iv) described above for two CHDM models with $\Omega_\nu = 0.2$ and 0.3, that even the model with $\Omega_\nu = 0.20$ fell a factor of two–three short of the Lanzetta *et al.* $z = 3 - 3.5$ data point. Unfortunately, the mass resolution in the simulation of MB94 is far too coarse to address the problem of the damped Ly$\alpha$ clouds. The mass of each of their "cold" particles was $2.5 \times 10^{10}$ $M_\odot$. In the simulations one would need to have at least 10-30 particles to be able to identify halos reliably. This already puts a limit of $(3 - 10) \times 10^{11}$ $M_\odot$ on possible collapsed halos in their simulations. This is a very large limit and is exactly the point where their curves $\Omega_g(M_{\text{halo}})$ stop growing in their Figure 2(a). Another important effect missing in the assumptions of MB94 is the cooling of the gas. The cooling time of the gas with the densities in the simulations is short. The observed gas is also cold (it is neutral gas by definition). Thus, there is no reason to assume as they do that the gas distribution follows that of dark matter. They also claim that the parameters of the Press-Schechter approximation for the CHDM model are different from what we find. We reanalyzed their results and find the same Press-Schechter parameters as for our simulations: Gaussian filter with $\delta_c = 1.5$ for halos with mean overdensity larger than 200. Figure 3 presents results of MB94 (circles connected by dashed curves). We indicate the number of cold particles per halo in the figure and show results for halos with insufficient number of particles (less than 100) with open circles. Results from the Press-Schechter approximation with Gaussian filter, $\delta_c = 1.50$, using the actual power spectrum in their simulations, correspond to the continuous curves. The Press-Schechter approximation gives reasonably accurate predictions for massive halos, which had more than 100 particles. Numerical results have a tendency to fall below the theoretical predictions once the number of particles per halo gets unreasonably small. Of course, one can find many interesting numerical effects once we get to the limit of 4 particles per halo, but it seems clear that the mass resolution is to blame for the problems, not the Press-Schechter approximation. Effects of the mass resolution are difficult to estimate. We tried to mimic them by cutting the spectrum of fluctuations at frequencies close to, but smaller than, the Nyquist frequency for particles. In Figure 4 (bottom panel) we show results for the Nyquist frequency cutoff (full curves) and for 1.4 smaller cutoff of the spectrum (long dashed curves). The cutoff has no effect on halos with large mass – just as one should expect because the Gaussian filter cuts off the power anyway. But for low mass objects the effect of the mass resolution becomes more significant – numerical results are closer to the theoretical predictions.

### 2.3 Finite Box Size and Mass Resolution Corrections

We note that linear corrections due to finite box size and mass resolution are usually not considered. They are almost negligable in the regime (at low $z$) when the Gaussian term in the Press-Schechter approximation is not too small (thus, when $n(z)$ curve is close to its top). But those corrections are very important for the epochs and masses when the Gaussian term is the leading one (on the steeply rising tail of $n(z)$). As an example, let's roughly estimate an error in the number of halos $n$, which would arise due to a small 10% error in rms fluctuations $\sigma$. For a typical situation at high $z$ one has $\sigma = 0.5$. Then for $\delta_c = 1.5$ we get the ratio $n_{\text{true}}/n_{\text{estimate}} = \exp\left((\delta_c/\sigma)(\Delta\sigma/\sigma)\right) = 2.5$ Thus, a 10% error in $\sigma$ resulted in the 150% error in $n$. The effect is very sensitive to the value of $\sigma$. Because $\sigma(M)$ decreases with mass, one expects that the correction is larger for larger masses. In our particular case of $20h^{-1}$Mpc box simulations, the error for $M = 10^{11}$ $M_\odot$ was about a factor of 1.5 (5% error in $\sigma$). Another error results in underestimate of the number of low mass halos. It is due to the lack of mass resolution: small objects will not appear because the initial spectrum did not have corresponding small waves. In this case $\sigma$ (the power predicted for this kind of objects) is typically larger ($\sigma \approx 1$), but the error in $\sigma$ is significantly larger (no waves, no $\sigma$). Results of MB94 were significuntly affected due to this lack of mass resolution.

Effects of the finite box size, the mass resolution, and galaxy indentification are illustrated in Figure 4 (top panel). The long dashed curves are the same as fulled curves on Figure 3: for 100 Mpc box, the mass reolution of MB94. If instead of taking the limits on the spectrum for the simulations, we take infinite box size and infinite resolution (of course formally only), we get the results shown as short-dashed curves. The corrections are not that big, but still at $z = 3$ the correction for $10^{11}$ $M_\odot$ is about a factor of two. If we change the parameter $\delta_c$ from 1.5 to 1.4 in order to recover some of the actually collapsed mass (as discussed above), the correction becomes even bigger. Nonlinear effects can probably raise the predictions even more. It is not unreasonable to extrapolate the parameter $\delta_c$ to an even lower value of, say, $\delta_c = 1.3$.

## 3. Damped Ly$\alpha$ Systems

If $\Omega_b$ is the baryon density parameter and $f_g$ is the neutral fraction of the gas in the absorbers, then the density parameter contributed by matter collapsed in structures associated with such systems is

$$\Omega_{coll} = \frac{\Omega_g}{\Omega_b f_g}. \qquad (7)$$

In the above relation, $\Omega_g$ is the total density parameter of neutral gas in damped Ly$\alpha$ systems, as provided by observational data (Lanzetta 1993, LWT93). In the following, we assume that $f_g = 1$, so that the neutral mass in a collapsed object of total mass $M$ is $\Omega_b M$. If CHDM cannot account for the damped Ly$\alpha$ observations with $f_g = 1$, it will have an even harder time if $f_g < 1$.

According to the Press-Schechter approximation, the fraction of the total volume, $F(M, z)$, associated with Gaussian–distributed fluctuations that collapse at redshift $z$ to structures of mass larger than $M$, is

$$F(M, z) = \text{erfc}\left(\frac{\delta_c}{\sqrt{2}\sigma(M, z)}\right), \qquad (8)$$

where erfc($x$) is the complementary error function. Under the assumption of very small fluctuations at the time the collapse starts, eq. (4) coincides with the contribution to the density parameter at redshift $z$ due to structures of mass $> M$.

In Figure 5 we plot the resulting $\Omega_{coll}$ as a function of redshift for three different fluctuation spectra and at different mass scales. Figure 5(a) refers to the old standard $\Omega_c/\Omega_\nu/\Omega_b = 0.6/0.3/0.1$ CHDM model while Figure 5(b) is for $0.675/0.25/0.075$ and Figure 5(c) is for $0.725/0.20/0.075$. This figure is to be compared with Figure 1 of Subramanian & Padmanabhan (1994). If we identify structures associated with Ly$\alpha$ absorbers as fluctuations with $\delta \geq \nu\sigma(r_f,z)$, then we will expect that, for fixed $\nu$, smaller scale fluctuations undergo non–linear evolution at higher redshifts. This is shown by the light dashed lines, which trace the redshifts at which fluctuations with $\nu = 1.5$ (upper line) and $\nu = 2$ (lower line) reach non–linearity (i.e., $\sigma(r_f,z) = \nu^{-1}$). For $M = 10^{12} M_\odot$ all the models severely underestimate $\Omega_{coll}$ at high redshifts, while taking $M = 10^{10} M_\odot$ is consistent with observations up to $z \simeq 3$. The highest-redshift point is not reproduced by the $0.6/0.3/0.1$ model, while the $0.675/0.25/0.075$ model seems only marginally inconsistent with it, and the $0.725/0.20/0.075$ model is clearly compatible with it.

We derive the relationship between the mass of a galaxy $M$, the comoving radius $r_f$ of the Gaussian filter, and the circular velocity $V_c$ in the following simplified way. For the spherical collapse model (e.g. Peebles 1980), the proper radius $r_g$ of a galaxy collapsed at redshift $z$ is $r_g = r_f \sqrt{2\pi}(1+z)^{-1}[(4\pi/3)178]^{-1/3}$. Then $V_c \equiv (GM/r_g)^{1/2}$ and we get

$$M = \frac{V_c^3}{\sqrt{89}GH_0(1+z)^{3/2}} = 2.45 \times 10^{11} \; h^{-1} M_\odot \left(\frac{V_c}{100 \text{ km s}^{-1}}\right)^3 (1+z)^{-3/2}. \quad (9)$$

This is the mass we insert in the Press-Schechter expression (1).

In Figure 6 we compare the predictions of two CHDM models (full curves) for the total density of baryons in damped Ly$\alpha$ systems with the observational data (Lanzetta 1993, LWT93). We again assume the neutral fraction is unity. The predictions always rise with decreasing $z$ and provide only upper limits to the mass in neutral hydrogen in collapsed objects of various $V_c$. We find that if most of the observed Ly$\alpha$ absorption originates from galaxies with circular velocity larger than 100 km $s^{-1}$, the standard $0.6/0.3/0.1$ CHDM model (left panel) has severe problems because it predicts too small a fraction of mass in these objects at redshifts $\geq 2.5$. However, if damped Ly$\alpha$ systems correspond to objects of circular velocity $\geq 50$ km $s^{-1}$, only the highest-redshift ($z = 3.0 - 3.4$) point is in serious disagreement with the predictions of the $0.6/0.3/0.1$ CHDM model. Because there is data in LWT93 on only four damped Ly$\alpha$ systems at $z > 3$ with no measurements of rotational velocities, it is difficult to say if the model is rejected or not. If, like quasars, the numbers of massive damped Ly$\alpha$ systems actually decline beyond $z = 3$ — for example, because the damped Ly$\alpha$ systems at higher redshift are mainly smaller-mass or still-collapsing systems (e.g., caustics) — then the model is still viable. But White, Kinney, and Becker (1993) present additional evidence that the number of damped Ly$\alpha$ systems continue to grow with redshift at least up to redshift 3.4. If this is true, then CHDM models with $\Omega_\nu \gtrsim 0.3$ are excluded.

The dot-dash line in Figure 6(a) shows the result of an analogous Press-Schechter calculation using a top-hat filter and $\delta_c = 1.68$. This is in excellent agreement with the $V_c > 50$ km s$^{-1}$ curve in Figure 1 of Kauffmann & Charlot (1994), calculated under the same assumptions. This explains why we disagree with the conclusions of this and the other recent preprints that used a Press-Schechter approximation with these assumptions: they correspond to numbers of halos considerably smaller at high redshifts than indicated by the high-resolution N-body simulations of KHPR93 and NKP94.

The prediction for the 0.6/0.3/0.1 CHDM model is that most of the gas at redshifts beyond about 2.5 is in small clustered clumps. Note that those clumps are not necessarily each protogalaxies. It is quite likely that numerous small clumps (at $z = 3$ they have dark matter masses typically $\approx 5 \times 10^8 \ h^{-1} M_\odot$, $V_c = 25$ km s$^{-1}$) can have a more massive neighbor — a forming galaxy, which could be observed at the redshift of the damped Ly$\alpha$ system. Later those clumps will merge and produce a real galaxy.

Predictions for numbers of objects at higher redshifts are exponentially sensitive to the fluctuation amplitude. Figure 6(b) shows that the $\Omega_c/\Omega_\nu/\Omega_b = 0.675/0.25/0.075$ variant of the CHDM model with a relatively small difference in parameters predicts significantly larger clumps. A typical object at $z = 3$ in this case has mass of $3.8 \times 10^9 \ h^{-1} M_\odot$, radius $6.6 h^{-1}$ kpc, and $V_c = 50$ km s$^{-1}$. This looks like a small (proto)galaxy. Now even the highest-$z$ point can be accounted for by halos with $V_c \sim 50$ km s$^{-1}$, certainly for $\Omega_\nu = 0.20$, and the question then arises what becomes of all this neutral hydrogen at lower redshifts.

We can make a more realistic (but more speculative) model for the damped Ly$\alpha$ systems if we suggest that it takes some time $\tau_{\rm gas}$ for each galaxy to "digest" the gas – to convert most the gas into stars or to ionize it. In order to mimic this situation we estimate as before the density of baryons confined at redshift $z(t)$ in halos with rotational velocity larger than $V_c$, but assume that *all* the baryons in such halos at a time $\tau_{\rm gas}$ earlier has now been ionized or converted to stars. Thus, we estimate the fraction of neutral gas at $z$ inside halos with $V > V_c$ is $\tilde{\Omega}_g(z) = \Omega_g(z) - \Omega_g(z(t - \tau_{\rm gas}))$, where $\tau_{\rm gas}$ is a characteristic life-time for the neutral gas. Like Kauffmann & Charlot (1994), we found that a constant $\tau_{\rm gas}$ cannot reproduce the observed trend in $\Omega_g$. We need to assume that this process of "digesting" was less efficient in the past. Results for $\tilde{\Omega}_g(z)$ for a model with $\tau_{\rm gas} = (1 + z)^2 0.35 \times 10^8$ yr are shown in Figure 6 as dashed curves.

### 4. Conclusions

The "standard" CHDM model with $\Omega_c/\Omega_\nu/\Omega_b = 0.6/0.3/0.1$ predicts a total density of neutral hydrogen in objects with mass $> 10^9 M_\odot$ and circular velocity $V_c > 50$ km s$^{-1}$ that is consistent with the damped Ly$\alpha$ system observations reported in LWT93 at $z \leq 3$, but not with their highest-redshift data point. However, these predictions are quite sensitive to the parameters of the model, and for example in CHDM with $\Omega_\nu = 0.20$ collapsed objects with $V_c \geq 50$ km s$^{-1}$ can account for all the data. Thus it appears to be premature to conclude that CHDM is inconsistent with the damped Ly$\alpha$ data.

The Table summarizes predictions of various CHDM and CDM models. Besides the CHDM models with 20, 25 and 30 per cent of the total mass in hot particles, we also show models with slightly different slope $n$ of the initial power spectrum (the superscript in model names indicates the slope). Three variants of the CDM model are also presented. All models except the COBE normalized CDM (CDMc) are normalized to the same amplitude on $8h^{-1}$Mpc scale. This roughly corresponds to the same number density of clusters. In detail, though, actual predictions for clusters are quite different even for the same $\sigma_8$. Klypin & Rhee (1994) and Walter & Klypin (1994) found that the number of rich ("Abell") clusters can be estimated using the Press-Schechter approximation with Gaussian filter and $\delta_c = 1.5$. Because Bryan et al. (1994) also found that the model with $\Omega_\nu = 0.30$ predicts a temperature distribution function of clusters that is compatible with observatioanl results, we use predictions for the model with $\Omega_\nu = 0.30$ as a guideline for the number of rich clusters. The model actually predicted a slightly larger number of very massive clusters, but the difference was not large. For comparison, the CDMc model predicts almost 5 times more rich clusters, which is widely recognized as a significant problem for the model. Biased CDM (CDMs) as well as tilted CDM (CDMt) predict too few rich clusters. The correlation length for the clusters in CDM models was too small. This is indicated by the small zero-crossing length $r_0$ of the correlation function. The tilt of the initial spectrum helps to solve this problem for CDMt, but it still does not cure the problem with the number of clusters. Furthermore, Plionis et al. (1994) showed that CDMt with $n = 0.7$ produces too weak clustering in the cluster distribution on scales $> 30\,h^{-1}$ Mpc. We expect the situation to be even worse for $n = 0.8$. Another statistic that probes large-scale power is the bulk velocity for a sphere of $50h^{-1}$Mpc radius. Observations indicate value $V(50\,h^{-1}$ Mpc$) \approx 335$ km s$^{-1}$ (Dekel 1994), which is a problem for both biased and tilted CDM. All CHDM variants predict enough large scale velocity, although variants that could be compatible with the Lanzetta et al. high-z data tend to predict $V_{50}$ which is systematically lower than that of the $\Omega_\nu = 0.3$ model. Finally, we show *linear* predictions for pair-wise velocities at $1h^{-1}$Mpc. This should not be compared with observed numbers because effects of nonlinearity are very significant. Nevertheless, the purpose of the values presented in the Table is to show that all CHDM models have significantly smaller power on $1h^{-1}$Mpc scale, which will be reflected in estimates of real pair-wise velocities.

It is very difficult to make a definite statement concerning pair-wise velocities. This is due to disarray in both observational estimates of the statistic and theoretical uncertainties due to very large cosmic variance.

It has long been assumed that the analysis of Davis & Peebles (1983) showing that the galaxy pairwise velocity dispersion $\sigma(1h^{-1}\text{Mpc}) \approx 320$ km s$^{-1}$ from the CfA1 redshift survey is a very tight constraint on cosmological models of structure formation, and indeed KHPR93 showed that 0.6/0/3/0.1 CHDM is consistent with this constraint. However, Mo, Jing, & Börner (1993) and Zurek et al. (1994) obtained considerably higher numbers for $\sigma(1\ h^{-1}$ Mpc) from various data sets, and also showed that $\sigma(r)$ is not a very robust statistic since it depends very strongly on the presence of rich clusters in the survey volume – these are both rare and hard to simulate accurately. Somerville, Davis, and Primack (in prep.) have independently confirmed these conclusions, and have also discovered an error in the code used by Davis & Peebles (1983) which is responsible for their low estimate of $\sigma(1\ h^{-1}$ Mpc).

It will be be necessary to check by running new N-body simulations that CHDM models with parameters such as 0.725/0.20/0.075 will still have velocities on small scales that are compatible with the data. The indications that $\sigma(1\ h^{-1}$ Mpc) are larger give us reason to expect that this will indeed be so. Moreover, the direct comparisons of CDM and 0.6/0.3/0.1 CHDM with the CfA data reported in NKP94 indicated that the small-scale velocities in these CHDM simulations were a little low, so there may be a little room available for lowering the neutrino mass (or, equivalently, $\Omega_\nu$). Much more detailed analysis of this sort, including the effects of breaking up overmerged halos (Nolthenius, Klypin, & Primack, in prep.), strengthens this conclusion. Finally, a preliminary analysis of results from a new N-body simulation of 0.725/0.20/0.075 CHDM with two equal-mass neutrinos suggests that small-scale velocities are sufficiently small to agree with the data (Primack et al. 1994). We found that 1D real-space pair-wise rms velocity for cold particles at $1\ h^{-1}$ Mpc separation is 670 km s$^{-1}$, which leads to a naive estimate of 540 km s$^{-1}$ for halos, if we take usual velocity bias $b_v = \sigma(\text{halos})/sigma(\text{dark matter})$. Projected pair-wise velocities can be estimated by placing an "observer" in the box and measuring relative velocities along line of sight for given projected separation. This gives projected pair-wise velocity at $1\ h^{-1}$ Mpc which is 20% smaller than in real space: $\sigma_v(\text{projected, darkmatter} = 560$ km s$^{-1}$. And assuming $b_v = 0.8$ we get an estimate for pair-wise velocities of halos: $\sigma_v = 450$ km s$^{-1}$. This is larger than the value found by Davis & Peebles, but is quite consistent with estimates found by Mo, Jing, & Börner (1993) and Zurek *et al.* (1994).

This paper also shows that COBE-normalized $\Omega_\nu = 0.20$ CHDM models do not overproduce clusters, especially if the neutrino mass is shared by two neutrinos, as suggested by some recent data. (Clusters are a concern, since it is well known that clusters are overproduced in the CDM limit $\Omega_\nu \longrightarrow 0$ if the fluctuation amplitude is normalized to COBE; cf. White et al. 1993a. If producing the observed numbers of clusters requires that the fluctuation amplitude be 10% lower than the central COBE value, for example, then we find that $\Omega_{coll}$ for $M = 10^{10} M_\odot$ decreases by a factor of about two at $z = 3$.) ¿From the analysis of cluster simulations based on the Zel'dovich approximation, we also find that the cluster 2-point correlation function for the $\Omega_\nu = 0.2$ CHDM model agrees with that measured for Abell/ACO clusters. This is also quite comfortable, since it is known that in the limit of a pure CDM model cluster correlations at scales $> 20 h^{-1}$Mpc are by far much weaker than measured in real data sets. A further test for CHDM models is also represented by the void probability function for the galaxy distribution. In a previous paper (Ghigna et al. 1994) we have shown that the $\Omega_\nu = 0.30$ CHDM model overproduces large voids, when compared to real data sets, while lowering $\Omega_\nu$ should reduce the presence of such large voids.

Of course, to investigate the formation of damped Ly$\alpha$ systems more fully will require not only such dissipationless simulations, but also inclusion of hydrodynamical effects. However, it is already clear that in a model such as CHDM, in which most large structures form at relatively low redshifts, most damped Ly$\alpha$ systems at redshifts beyond about 2.5 must be associated with objects having relatively small circular velocities $V_c < 100-150$ km $s^{-1}$. Although it is impossible to estimate the circular velocities or velocity dispersions of such systems by observing the Ly$\alpha$ absorption itself (since its width is due to damping) such systems typically have substantial metallicity. The Doppler widths of the associated metal line systems or the relative velocities of different metal-rich clouds, which can be measured for example using the high-resolution spectrograph at the Keck telescope, should allow estimates of the velocities, and therefore the masses, of these objects. The clear prediction of CHDM is that these widths, relative velocities, and masses must decline with increasing redshift.

**Acknowledgments.** SB thanks UCSC for its hospitality during the first phase of preparation of this work. JRP acknowledges very helpful conversations with D. N. Schramm, K. Lanzetta, T. Walker, and A. Wolfe, and support from NSF and the University of California, Santa Cruz. Simulations were done using the CONVEX-3880 at the NCSA.

FIGURE CAPTIONS

**Figure 1**. (a) Power spectra of fluctuations in the total density at the present epoch for CHDM and CDM with $\Omega = 1$. The lower three curves are for CHDM models with various fractions of baryons and hot dark matter (neutrinos). The lower full curve is for $\Omega_c/\Omega_\nu/\Omega_b = 0.60/0.30/0.10$. The dashed curve is for $0.655/0.27/0.075$. The dot-dashed curve is for $0.675/0.25/0.075$. The upper full curve is for the CDM model with the same normalization on very long waves.

(b) Number density of objects at different redshifts in the $0.60/0.30/0.10$ CHDM model. Results of numerical simulations are shown as dashed (box size $25h^{-1}$Mpc) and full (box size $50h^{-1}$Mpc) curves. The dot-dashed curves indicate predictions based on the Press-Schechter approximation. The value of $\delta_c$ for a Gaussian filter for each mass is indicated in the Figure. The dotted curve is for a Gaussian filter with $\delta_c = 1.33$. Triangles show predictions for the top-hat filter with $\delta_c = 1.68$.

**Figure 2**. Number of halos in the CHDM model with $\Omega_\nu = 0.25, \Omega_b = 0.075$. The resolution in this case was $25h^{-1}$kpc in comoving coordinates or $6.2h^{-1}$kpc in proper units at $z = 3$, the box size was $20h^{-1}$Mpc. The top panel shows the evolution of the number density of halos defined as having the mean overdensity above 200. The bottom panel compares numerical results at $z = 3.2$ with the Press-Schechter approximation. The lower triangles are the predictions of the Press-Schechter approximation with the Gaussian filter, $\delta_c = 1.5$, and the power spectrum actually considered in the simulation – only waves shorter than the box size and longer than the Nyquist frequency of particles were considered. The full curve is for the halos with overdensity 200. The upper triangles are for $\delta_c = 1.40$. The dot-dashed curve corresponds to halos, whose central overdensity is larger than 200, but whose mass is estimated at the overdensity 100. In this case the algoritm find the same halos as before, but assigns larger masses to them. The dashed curve is for halos with the mean overdensity 100 with additional constraint on the radius of halos: the proper radius should be less than 43 kpc ($h = 0.5$), which corresponds to comoving radius 180 kpc. Small halos were the same as found with the additional check of the central overdensity. Deviations were found only for very small halos (less than $2 \times 10^9$ $h^{-1}M_\odot$ ). This means that all our halos with the mean overdensity 100 had central overdensities larger than 200. The parameter for the Press-Schechter aprroximation was $\delta_c = 1.4$ for two last prescriptions.

**Figure 3**. Number of halos in Ma & Bertchinger (1994) simulations (circles connected by dashed curves). Results for halos with insufficient number of particles (less than 100) are shown with open circles. Results from the Press-Schechter approximation with Gaussian

filter, $\delta_c = 1.50$, and the actual power spectrum in their simulations are presented by full curves. The Press-Schechter approximation gives reasonably accurate predictions for massive halos, which had more than 100 particles. Numerical results have a tendency to fall below the theoretical predictions once the number of particles per halo gets unreaonably small.

**Figure 4.** Effects of the mass and finite box size on the predictions of the number of halos. The bottom panel shows results of Ma & Berchinger (1994) simulations for 100 Mpc ($h = 0.5$) box and $128^3$ cold particles. The Press-Schechter approximation for the same box and for wavelengths longer than the Nyquist frequency cutoff are shown as full curves. Long dashed curves are for waves 1.4 times longer the Nyquist cutoff. The cutoffs has no effect on halos with large mass. But for low mass objects the effect of the mass resolution became more significant – numerical results are closer to the theoretical predictions. The top panel illustrates combined effects of the finite box size the the mass resolution. The long dashed curves are the same as fulled curves on Figure 3: for 100 Mpc box, the mass reolution of Ma & Bertchinger. If instead of taking the limits on the spectrum for the simulations, we take infinite box size and infinite resolution, we get the results shown as short-dashed curves. The corrections at $z = 3$ for $10^{11}$ $M_\odot$ is about a factor of two. If we change the parameter $\delta_c$ from 1.5 to 1.4 in order to recover some of actually collapsed mass, the correction becomes even bigger. Actual nonlinear effects can probably rise the predictions even more. It is not unreasonable to extrapolate the paramer $\delta_c$ to even lower value of, say $\delta_c = 1.3$.

**Figure 5.** Fraction of the total density contributed by collapsed objects as a function of redshift. The data points are obtained from the observational data by Lanzetta et al. (1993) after suitably rescaling the total density in form of Ly$\alpha$ clouds (see text); indicated error bars correspond to $1\sigma$. Panel (a) is for the CHDM model with $\Omega_c/\Omega_\nu/\Omega_b = 0.60/0.30/0.10$, panel (b) is for $0.675/0.25/0.075$ and panel (c) is for $0.725/0.020/0.075$. The light dotted lines trace the redshifts at which fluctuations of $1.5\sigma(r_f, z)$ and $2\sigma(r_f, z)$ (upper and lower lines, respectively) reach non–linearity. Continuous, long-dashed and short-dashed curves correspond to $10^{10}$, $10^{11}$ and $10^{12}$ for the smallest mass of the collapsed structures. In all the cases, a Gaussian window is assumed with critical density contrast $\delta_c = 1.33$.

**Figure 6.** Fraction of the total density in the form of damped Ly$\alpha$ clouds: observational results of Lanzetta et al. (1993) compared with predictions of CHDM models with $\Omega_c/\Omega_\nu/\Omega_b$ given by (a) $0.60/0.30/0.10$, and (b) $0.675/0.25/0.075$. Error bars for the observational points correspond to $1\sigma$. The full curves indicate the fraction of baryons in collapsed objects with estimated circular velocities larger than $15, 25, 50, 100$ km s$^{-1}$ (from top to bottom), calculated using a Gaussian filter with $\delta_c = 1.33$. The curve for 15 km s$^{-1}$ is not shown on (b). The Press-Schechter predictions for a top-hat filter with $\delta_c = 1.68$ and circular velocity 50 km $s^{-1}$ are shown as the dot-dashed curve in (a). Results for a model of conversion of gas to stars with $\tau_{\rm gas} = (1+z)^2 0.35 \times 10^8$ y are shown as dashed curves (see text).

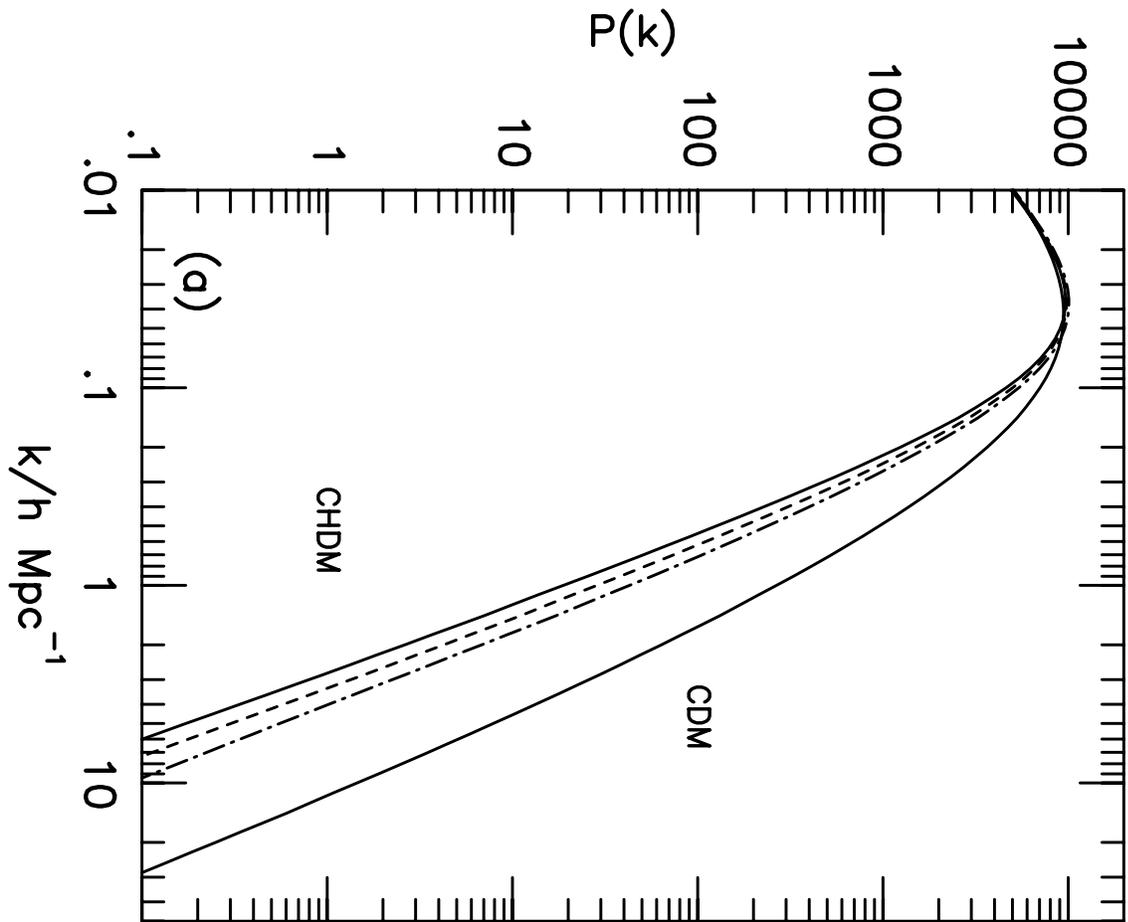

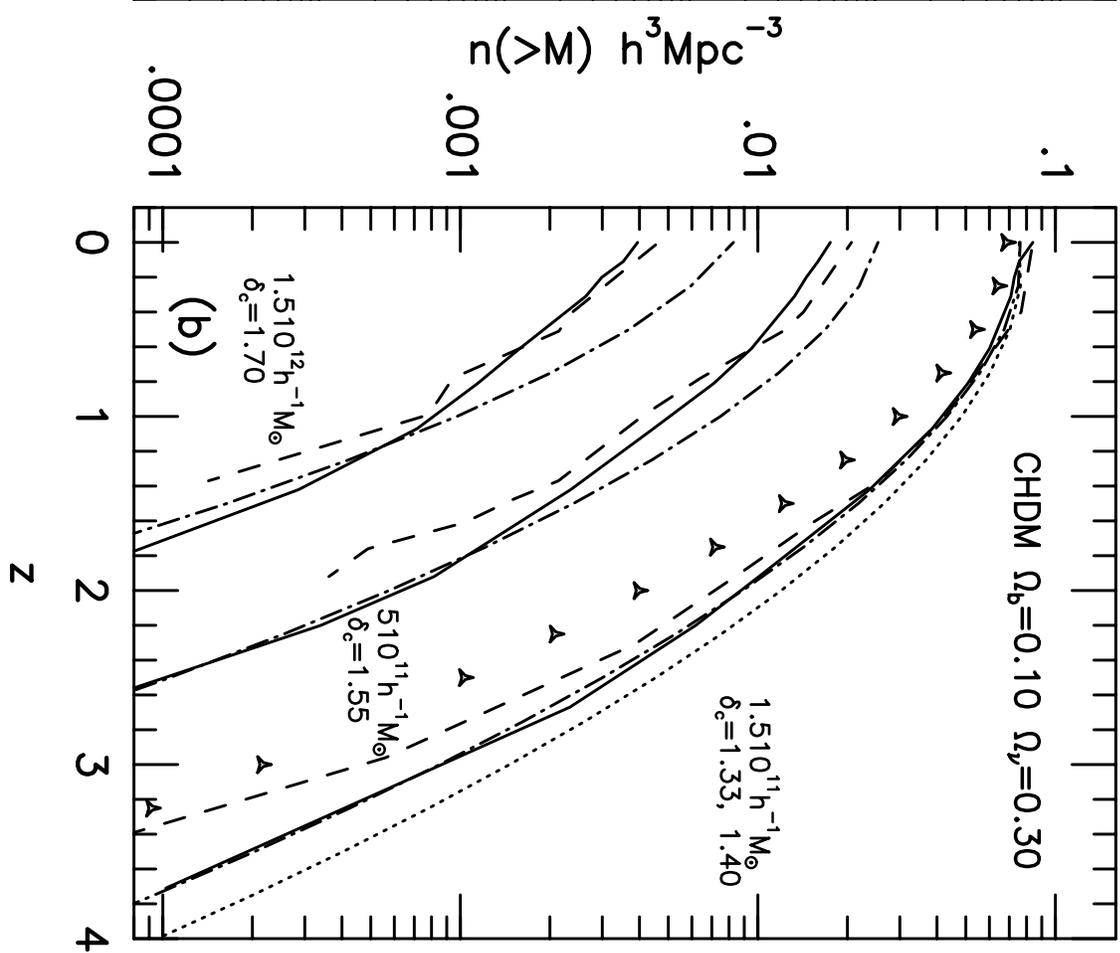

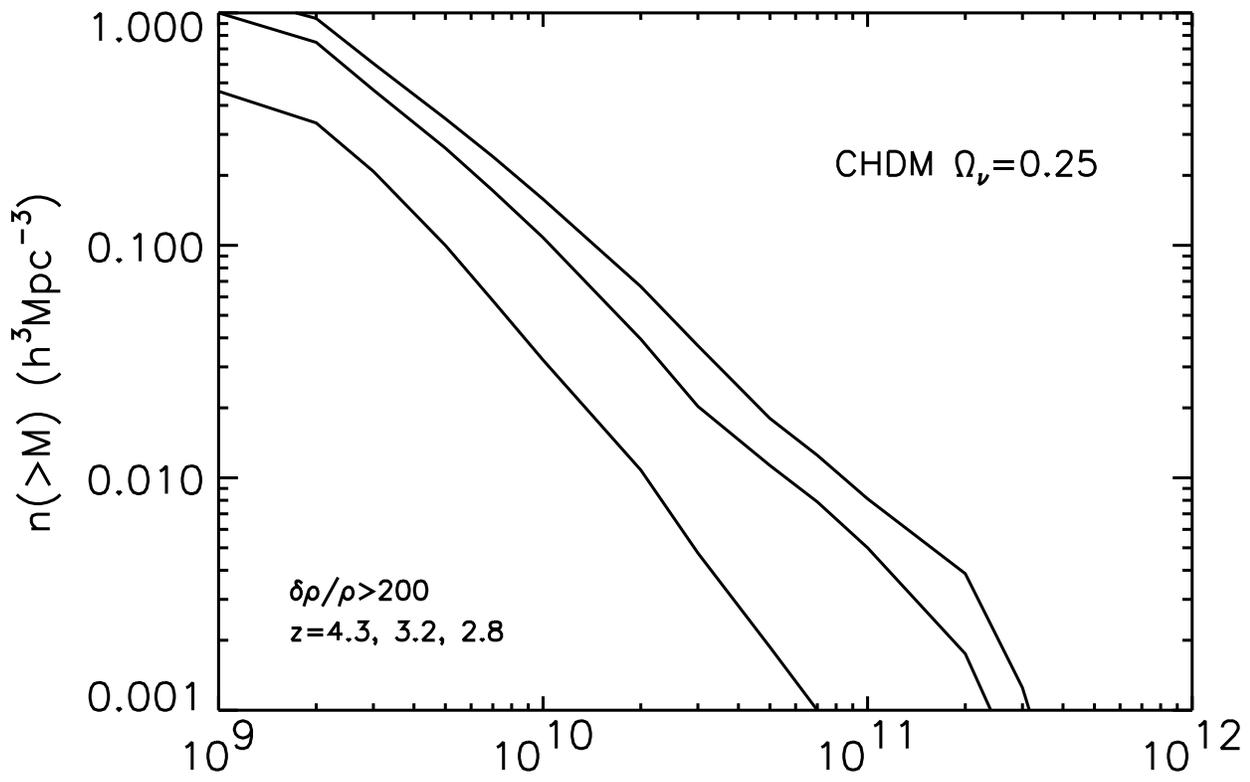
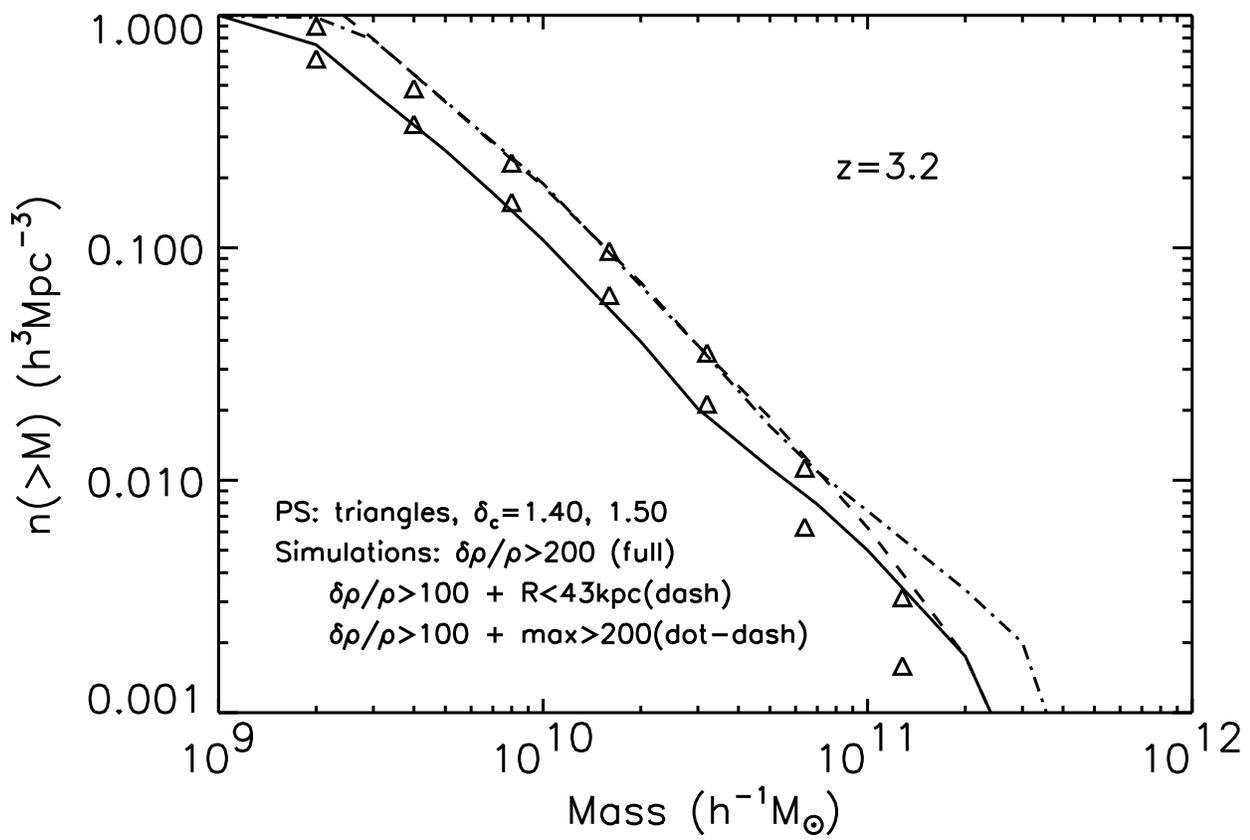

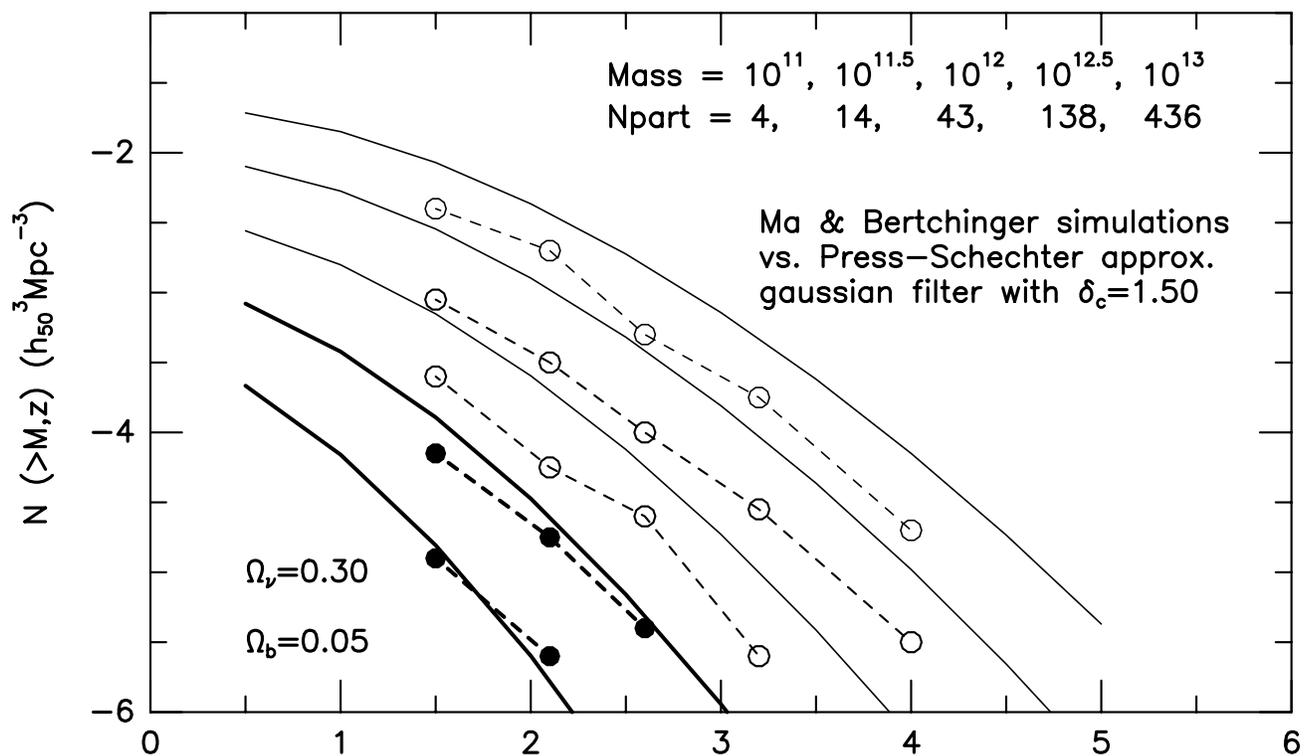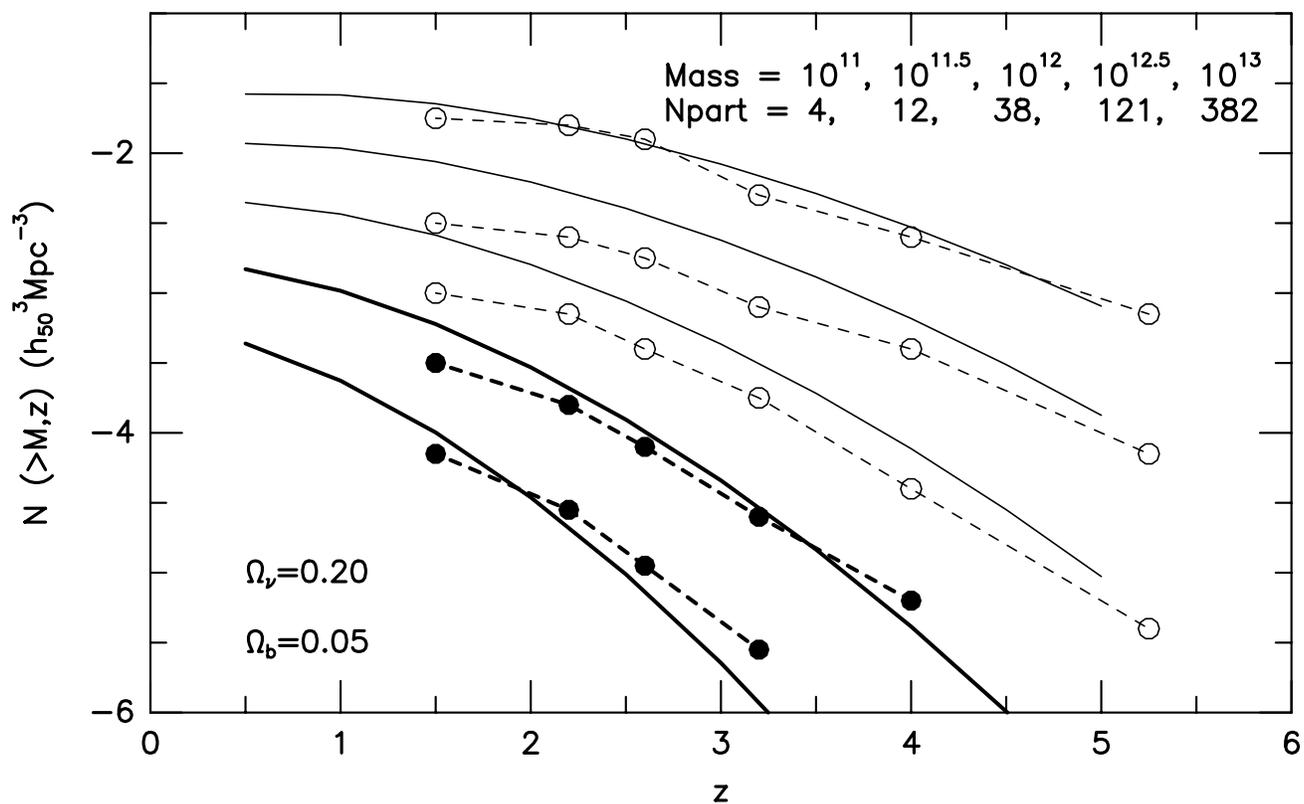

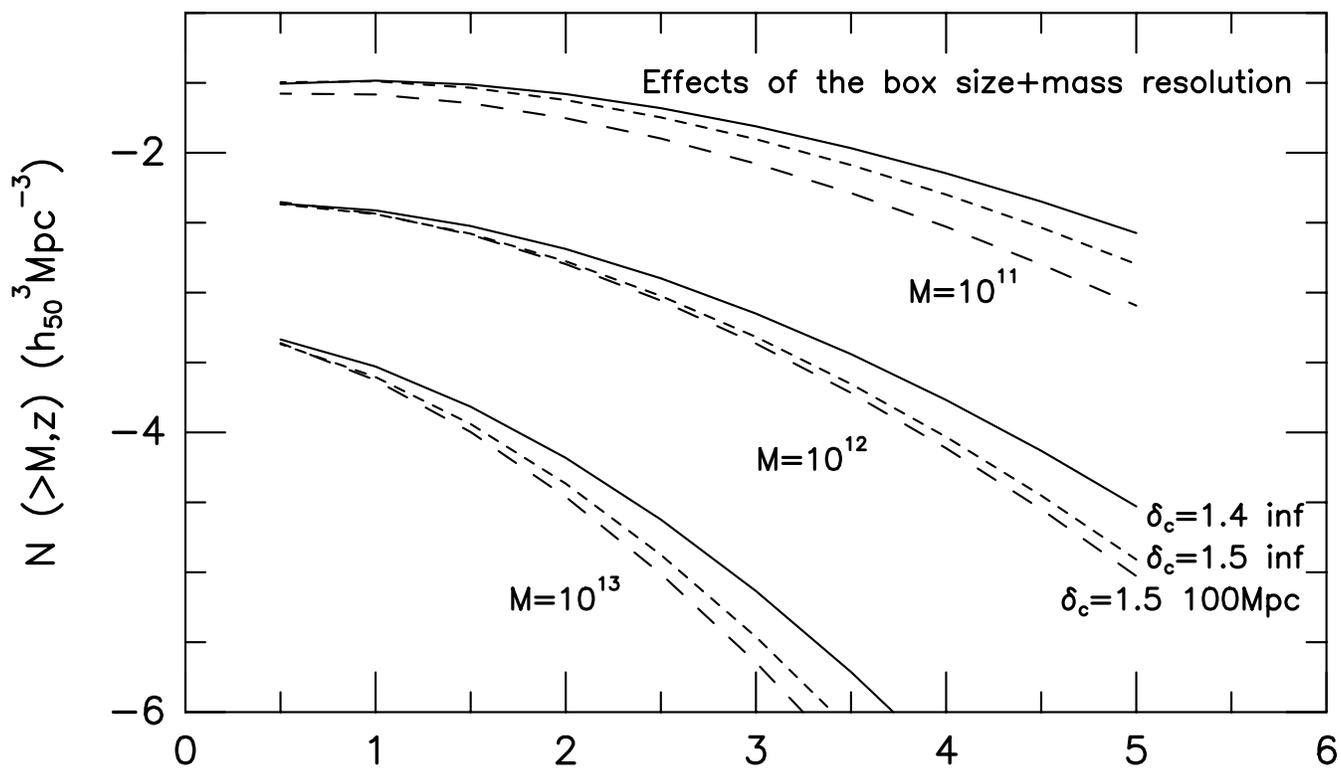
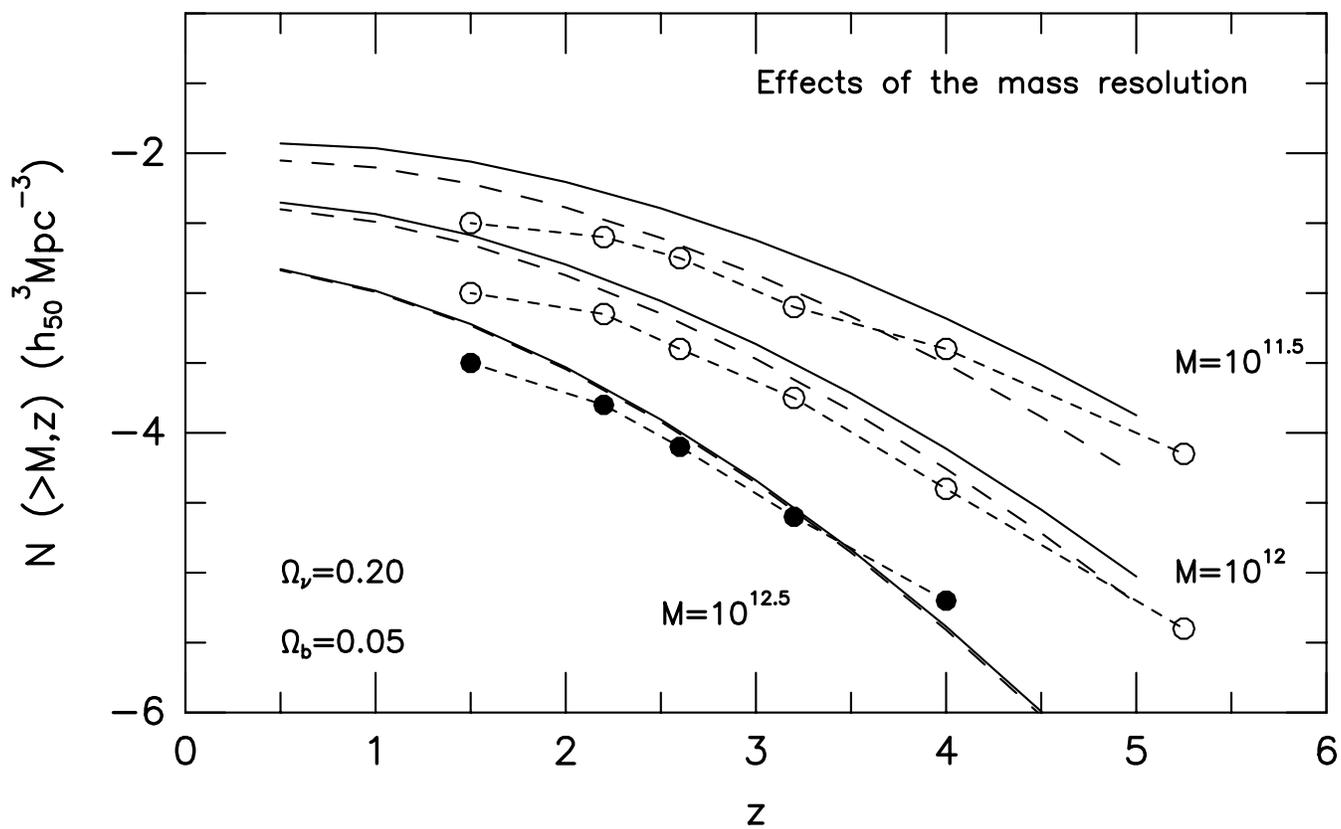

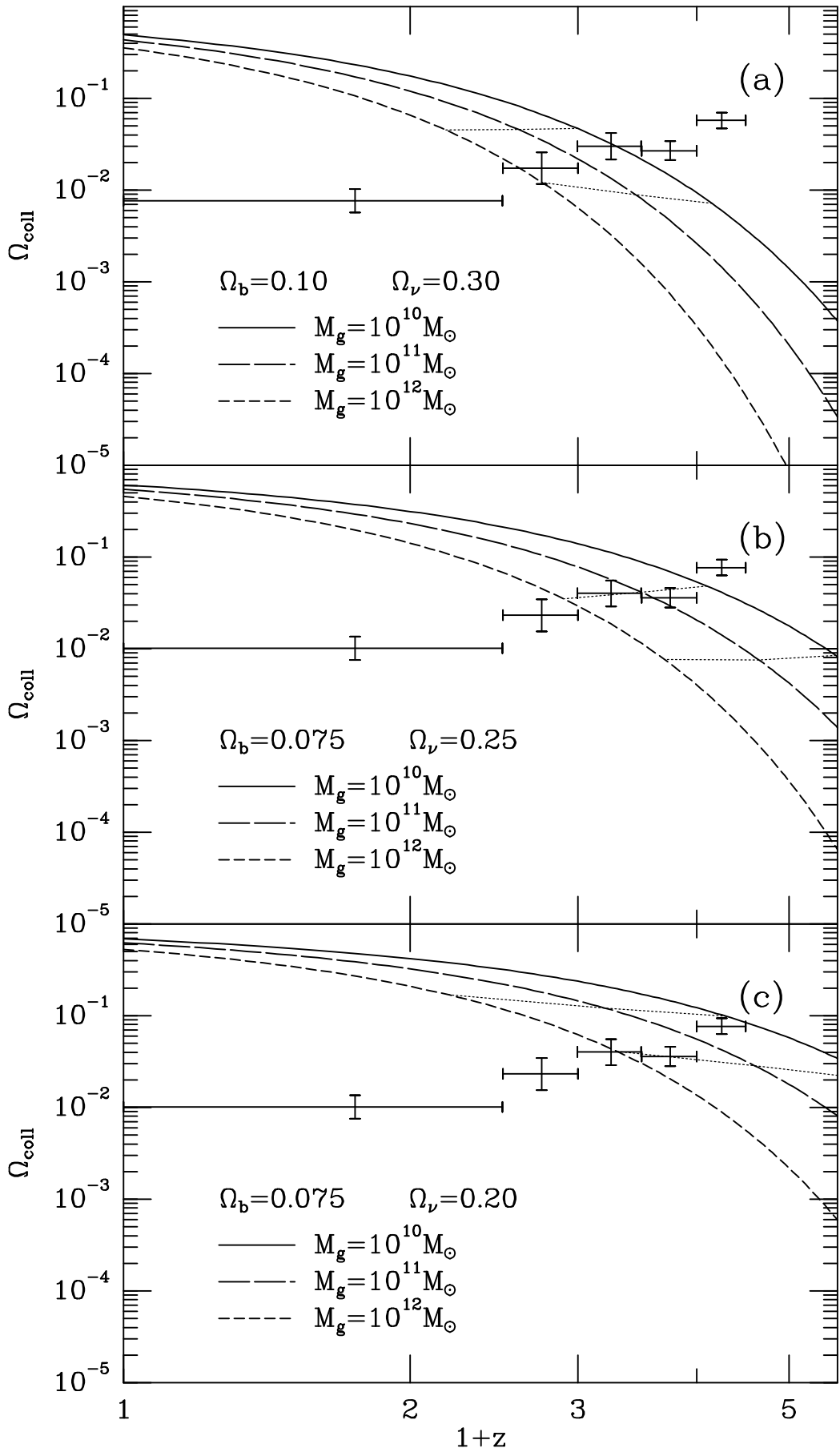

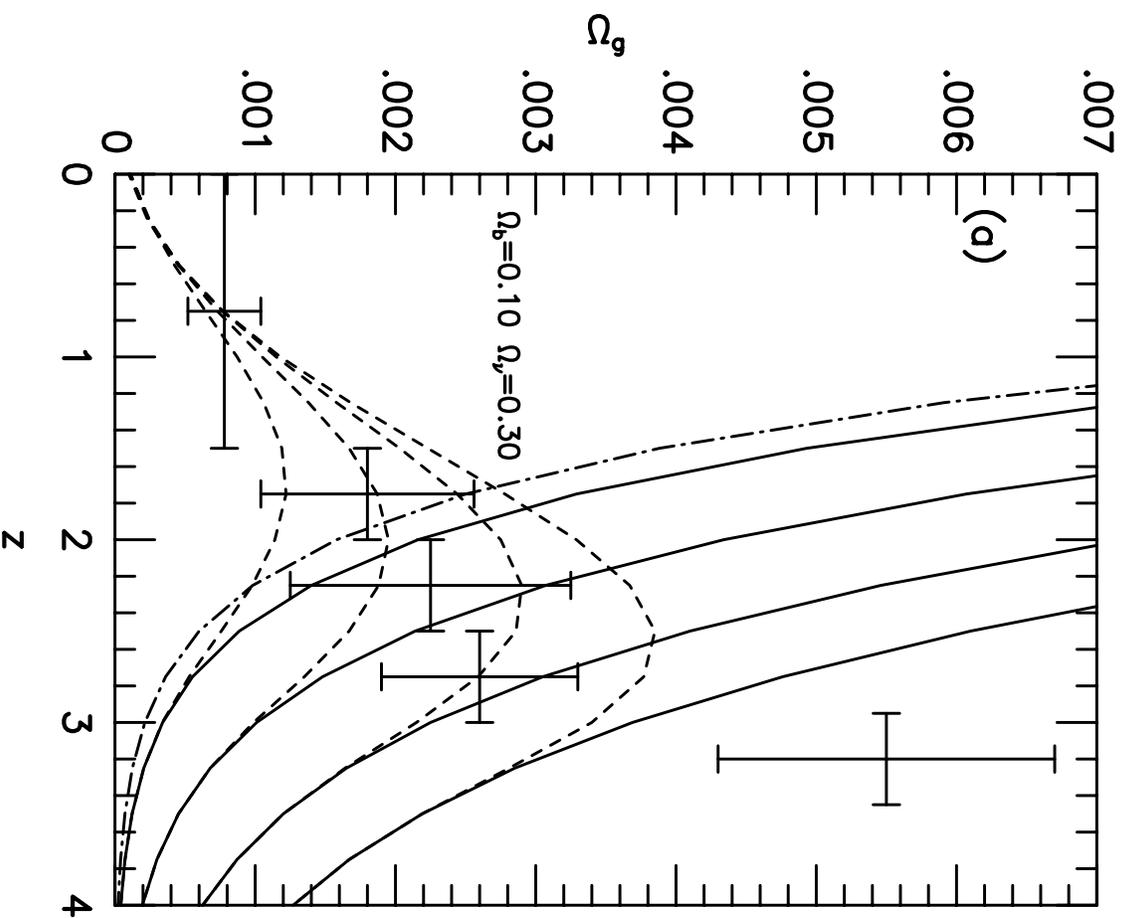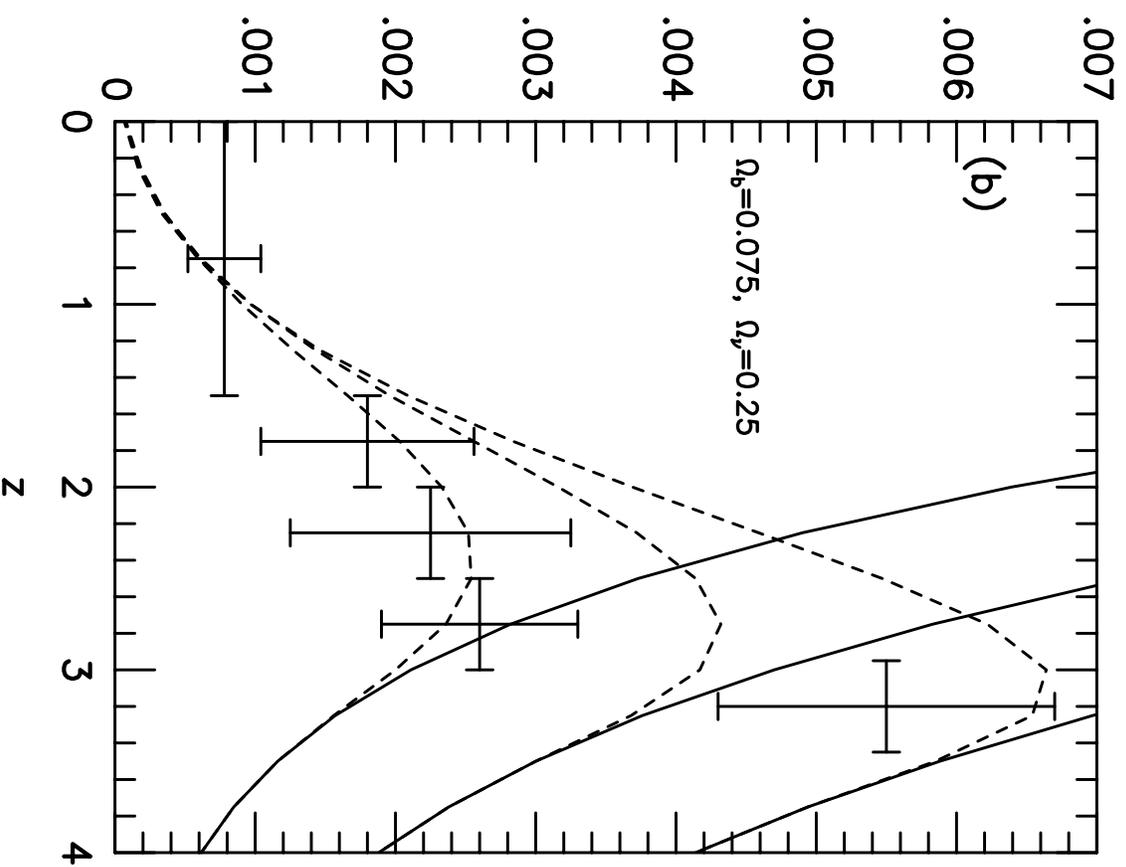

TABLE

PARAMETERS OF MODELS

| Parameter | | $M_{30}^1$ | $M_{30}^{1.1}$ | $M_{30}^{0.9}$ | $M_{20}^1$ | $M_{20}^{1.1}$ | $M_{20}^{0.9}$ | $M_{25}^1$ | CDMs | CDMc | CDMt |
|---|---|---|---|---|---|---|---|---|---|---|---|
| $\Omega_\nu$ | | 0.30 | 0.30 | 0.30 | 0.20 | 0.20 | 0.20 | 0.25 | 0 | 0 | 0 |
| $\Omega_{\text{baryon}}$ | | 0.10 | 0.10 | 0.10 | 0.075 | 0.075 | 0.075 | 0.075 | 0.075 | 0.075 | 0.075 |
| $n$ | a | 1.00 | 1.10 | 0.90 | 1.0 | 1.1 | 0.9 | 1.0 | 1.0 | 1.0 | 0.8 |
| $\sigma_8$ | b | 0.70 | 0.70 | 0.70 | 0.70 | 0.7 | 0.7 | 0.7 | 0.7 | 1.0 | 0.7 |
| $Q_{2,\text{rms}}$ | c | 17.87 | 13.48 | 23.70 | 15.37 | 11.53 | 20.50 | 16.23 | 11.61 | 16.58 | 21.0 |
| $\Delta T/T, l=3\text{--}10$ | d | 30.37 | 23.94 | 38.57 | 26.12 | 20.48 | 33.36 | 27.60 | 19.73 | 28.18 | 32.7 |
| $n(>10^{15}M_\odot)$ | e | 6.5e-7 | 5.8e-7 | 7.4e-7 | 3.9e-7 | 3.4e-7 | 4.6e-7 | 5.0e-7 | 1.1e-7 | 3.0e-6 | 1.6e-7 |
| $n(>2\cdot 10^{14}M_\odot)$ | e | 2.8e-5 | 2.9e-5 | 2.7e-5 | 3.1e-5 | 3.3e-5 | 3.1e-5 | 3.0e-5 | 3.9e-5 | 1.1e-4 | 3.6e-5 |
| $V_{50\text{ Mpc}}$ | f | 380 | 347 | 416 | 330 | 300 | 365 | 349 | 218 | 312 | 270 |
| $r_0$ | g | 53.5 | 50.0 | 57.5 | 51.0 | 47.0 | 55.5 | 51.5 | 36.0 | 36.0 | 46 |
| $\sigma_{M=10^{11}, z=3}$ | h | 0.442 | 0.480 | 0.409 | 0.598 | 0.656 | 0.546 | 0.520 | 1.06 | 1.52 | 0.873 |
| $\sigma_{M=10^{11}, z=4}$ | h | 0.358 | 0.389 | 0.331 | 0.483 | 0.530 | 0.441 | 0.420 | 0.866 | 1.24 | 0.71 |
| $\sigma_v(1h^{-1}\text{Mpc})$ | i | 104 | 114 | 96 | 146 | 163 | 133 | 125 | 280 | 400 | 221 |
| $M_{\text{halo}}$ $\delta_c$ $\Omega_g$ | | | | | | | | | | | |
| $2\cdot 10^{10}$ 1.33 | j | 9.7e-4 | 2.1e-3 | 4.0e-4 | 6.1e-3 | 9.6e-3 | 3.5e-3 | 2.8e-3 | 2.7e-2 | 4.0e-2 | 3.4e-2 |
| $2\cdot 10^{10}$ 1.40 | j | 6.3e-4 | 1.5e-3 | 2.4e-4 | 4.8e-3 | 8.0e-3 | 2.7e-3 | 2.0e-3 | 2.5e-3 | 3.9e-2 | 3.2e-2 |
| $1\cdot 10^{11}$ 1.33 | j | 3.8e-4 | 8.5e-4 | 1.6e-4 | 2.9e-3 | 4.8e-3 | 1.6e-3 | 1.2e-3 | 1.9e-2 | 3.2e-2 | 2.6e-2 |
| $1\cdot 10^{11}$ 1.40 | j | 2.3e-4 | 5.5e-4 | 8.6e-5 | 2.1e-3 | 3.8e-3 | 1.1e-3 | 8.4e-4 | 1.7e-2 | 3.0e-2 | 2.4e-2 |

a Slope of the power spectrum: $P(k) = Ak^n$

b $\Delta M/M$ for top-hat filter with $R_{\text{top-hat}} = 8h^{-1}\text{Mpc}$

c Rms value for the quadrupole component of $\Delta T$ in units of $\mu K$. The amplitude of the spectrum $A$ and the quadrupole of $\Delta T/T$ are related to each other: $A = Q_2^2 R_H^{n+3} \frac{\pi^2}{5}(3-\epsilon)(2-\epsilon)3^{1-n}\exp(1.4736\epsilon - 0.3612\epsilon^2)$, where $R_H = 6000/h$ Mpc is the distance to the horizon and $\epsilon = (n-1)/2$. The approximation gives four digits for $|\epsilon| < 0.1$.

d $\sqrt{\sum_3^{10} Q_l^2 \exp(-l(l+1)\sigma_a^2)}$ is the contribution of multipoles from $l = 3$ to $l = 10$ to the rms fluctuation of $\Delta T/T$ measured with antenna with gaussian width $\sigma_a = \text{FWHM}/2.355$, FWHM = $6°$.

$$Q_l^2 = \frac{A}{R_H^{n+3}} \frac{2l+1}{4\pi} \frac{\Gamma(3-n)\Gamma(l-1/2+n/2)}{2^{1-n}\Gamma^2(2-n/2)\Gamma(l+5/2-n/2)}$$

$$\approx Q_2^2 \frac{2l+1}{5} \frac{(3-\epsilon)(2-\epsilon)}{(l+1-\epsilon)(l-\epsilon)} \left(\frac{l}{2}\right)^{2\epsilon} \exp\left(-\epsilon[\frac{1}{l} + \frac{1}{6l^2} - \frac{13}{24}]\right)$$

e Number density of clusters above the mass limit in units of $(h^{-1}\text{ Mpc})^{-3}$. The Press-Schechter approximation with gaussian filter and $\delta_c = 1.50$ was used to make the prediction.

f Bulk velocity (km/sec) for a sphere of $50h^{-1}\text{Mpc}$ radius

g Zero crossing in $h^{-1}\text{Mpc}$ of the correlation function: $\xi(r_0) = 0$.

h $\Delta M/M$ at given redshift for gaussian smoothing with mass $10^{11}h^{-1}M_\odot$.

i Linear estimate of pair-wise velocities on $1h^{-1}\text{Mpc}$ scale.

j Mean density of the cold gas at $z = 3\text{--}3.5$ in objects with mass larger then the limit ($M_{\text{halo}}$ is in units of solar mass, $h = 0.5$, gausian filter)